\documentclass[letterpaper, 10 pt, conference]{ieeeconf}
\IEEEoverridecommandlockouts                              %

\usepackage[english]{babel}
\addto\captionsenglish{\renewcommand{\figurename}{Fig.}}

\PassOptionsToPackage{utf8}{inputenc}
\usepackage{inputenc}

\PassOptionsToPackage{T1}{fontenc}
\usepackage{fontenc}

\usepackage{cite}
\usepackage{amssymb,amsfonts,amsthm}
\usepackage{algorithmic}
\usepackage{graphicx}
\usepackage{textcomp}
\usepackage{xcolor}
\usepackage{xspace}
\usepackage[size=tiny]{todonotes}
\renewcommand{\todo}[1]{}
\usepackage{balance}
\usepackage[all]{nowidow}
\PassOptionsToPackage{fleqn}{amsmath}       %
\usepackage{amsmath}
\let\labelindent\relax
\usepackage[inline]{enumitem}
\usepackage[vlined,ruled,linesnumbered,longend]{algorithm2e}
\usepackage{svg}

\usepackage{booktabs}
\usepackage{siunitx}
\usepackage{tabularx}
\usepackage{adjustbox}
\usepackage{multirow}

\usepackage[hidelinks]{hyperref}
\usepackage{cleveref}
\Crefformat{figure}{#2Fig.~#1#3}
\Crefmultiformat{figure}{Figs.~#2#1#3}{ and~#2#1#3}{, #2#1#3}{ and~#2#1#3}
\usepackage{subfig}
\newsavebox{\measurebox}

\newcommand{\ie}{i.\,e.\xspace}

\newcommand{\eg}{e.\,g.\xspace}

\def\quad{\hskip1em\relax}
\usepackage{bm}
\newtheorem{example}{Example}
\newtheoremstyle{example}
{1pt}   %
{1pt}   %
{\itshape} %
{}      %
{\bfseries} %
{.}     %
{ }     %
{}      %

\renewcommand{\baselinestretch}{0.94}

\title{A Fully Planar Approach to Field-coupled Nanocomputing: Scalable Placement and Routing Without Wire Crossings}

\author{Benjamin Hien, Marcel Walter, Simon Hofmann, and Robert Wille%
	\thanks{Benjamin Hien, Marcel Walter, Simon Hofmann, and Robert Wille are with the Chair for Design Automation, Technical University of Munich, Germany. Marcel Walter and Robert Wille are also with the Munich Quantum Software Company GmbH, Garching near Munich, Germany. Robert Wille is also with the Software Competence Center Hagenberg GmbH (SCCH), Austria.
		E-mail: \{\href{mailto:benjamin.hien@tum.de}{\mbox{benjamin.hien}}, \href{mailto:marcel.walter@tum.de}{\mbox{marcel.walter}}, \href{mailto:simon.t.hofmann@tum.de}{\mbox{simon.t.hofmann}}, \href{mailto:robert.wille@tum.de}{\mbox{robert.wille}}\}@tum.de%
	}
	\\[0.5ex]
	\url{https://www.cda.cit.tum.de/research/nanotech/}
}

\begin{document}

\maketitle

\begin{abstract}
	\emph{Field-coupled Nanocomputing} (FCN) is a class of promising post-CMOS technologies that transmit information through electric or magnetic fields instead of current flow. They utilize basic building blocks called \emph{cells}, which can form gates that implement Boolean functions. However, the design constraints for FCN circuits differ significantly from those for CMOS. One major challenge is that wires in FCN have to be realized as gates, \ie, they are constructed from cells and incur the same costs as gates. Additionally, all FCN technologies are fabricated on a single layer, \eg, a silicon surface, requiring all elements---gates and wires---to be placed within that same layer. Consequently, FCN employs special gates, called wire crossings, to enable signals to cross. While existing wire-crossing implementations are complex and were previously considered costly, initial efforts have aimed at minimizing their use. However, recent physical simulations and experiments on a quantum annealing platform have shown that currently used wire crossings in FCN significantly compromise signal stability, to the extent that circuits cannot function reliably. This work addresses that issue by introducing the first placement and routing algorithm that produces fully planar FCN circuits, eliminating the need for all wire crossings. For a comparative evaluation, a state-of-the-art placement and routing algorithm was also modified to enforce planarity. However, our proposed algorithm is more scalable and can handle inputs with up to~$149k$ gates, enabling it to process circuits that are~$182 \times$ more complex than those handled by the modified state-of-the-art algorithm.
\end{abstract}

\section{Introduction \& Motivation}\label{sec:intro}

\emph{Field-coupled Nanocomputing} (FCN, \cite{Anderson14}) encompasses emerging nanotechnologies that depart from CMOS by utilizing physical fields instead of transistors and current flow for computation. This signal transfer mechanism allows FCN to operate near the \emph{Landauer Limit}\cite{landauer1961irreversibility}, offering a highly energy-efficient technology for modern computing. Recent breakthroughs in FCN fabrication, particularly via \emph{Silicon Dangling Bonds} (SiDBs, \cite{pitters2024atomically}), enable logic devices and interconnects smaller than $\SI{30}{\square\nano\meter}$\cite{huff2018binary}, driving progress in the field. SiDBs also facilitate \emph{Quantum-dot Cellular Automata} (QCA, \cite{lent2003molecular}), a foundational technology for numerous studies~\cite{hofmann2024wiring, hofmann2023post, hofmann2024born}.

Despite differing physical implementations, FCN technologies share similar design constraints in logic synthesis and physical design, allowing layouts for QCA and SiDB, at an abstract level, to be translated into one another~\cite{hofmann2023hexagonalization}. However, these constraints differ significantly from CMOS. A key distinction is the high cost of wires in FCN, which, unlike in CMOS, are as expensive as gates since both are built from elementary cells~\cite{reis2016methodology}. Hence, while the design process in CMOS commonly focuses on reducing delay, area, and power consumption~\cite{mishchenko2006dag}, in FCN, design constraints like minimizing wiring effort are much more emphasized.

Additionally, FCN is limited to a single layout layer for both gates and wiring, whereas CMOS utilizes multiple metal layers for routing without signal integrity issues. Consequently, FCN relies on wire crossings---specialized gate types that enable signals to cross on the same layer---which are standard components in modern \emph{Placement and Routing} (P\&R) algorithms.

Since wire crossing implementations are physically complex, they were initially regarded as costly. Consequently, research focused on crossing-aware design for FCN technologies to reduce this cost factor, while balancing the significant area overhead associated with layouts that minimize crossing counts\cite{marakkalage2024technology, hien2024reducing}. These studies primarily targeted wire crossings during the logic synthesis process, although they did not fully eliminate them. Moreover, these methods have not been validated at the physical design stage due to the absence of a comprehensive design process, particularly lacking a crossing-aware P\&R algorithm.

However, recent research on SiDB crossings~\cite{drewniok2023temperatureIEEE}, along with earlier studies on QCA crossings~\cite{retallick2017embeddingquantum}, has shown that wire crossings are ultimately impractical to implement in FCN. Thus, state-of-the-art P\&R algorithms are categorically inapplicable. This emphasizes the need for P\&R algorithms that avoid the use of wire crossings.

This work proposes a fully planar approach that eliminates wire crossings, addressing key challenges in FCN. To achieve this, we introduce a novel design flow for planar FCN circuits, featuring our key contribution: a scalable planar P\&R algorithm. First, we show that while a state-of-the-art P\&R method~\cite{hofmann2024born} can be adapted to generate planar layouts for small-scale networks, it lacks scalability and is insufficient for real-world applications. To overcome this, we present a new scalable P\&R algorithm that eliminates wire crossings and processes circuits with up to~$149k$ gates---$182\times$ larger than previous methods---paving the way for FCN to become a viable technology.

This paper is structured as follows: Section~\ref{sec:back} overviews FCN technologies, while Section~\ref{sec:related} reviews recent work on wire crossings. Section~\ref{sec:proposed} introduces our main contribution---a scalable, planar P\&R algorithm. Section~\ref{sec:experiments} presents experimental evaluations, comparing its performance to a modified P\&R algorithm. Finally, Section~\ref{sec:conclusion} concludes the paper.

\section{Background}\label{sec:back}

This section reviews the foundational concepts essential for FCN technologies. Section~\ref{sec:back:cell} details the primary components of FCN, while Section~\ref{sec:back:clock} explores clocking mechanisms within FCN frameworks. %

\begin{figure}[!t]
	\centering
	\subfloat[Basic FCN cells.]{\includegraphics[width=0.75\linewidth]{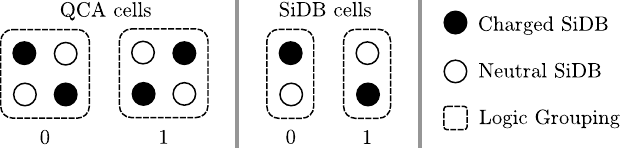}\label{fig:cells}}\\
	\subfloat[QCA MAJ gate.]{\includegraphics[height=0.27\linewidth]{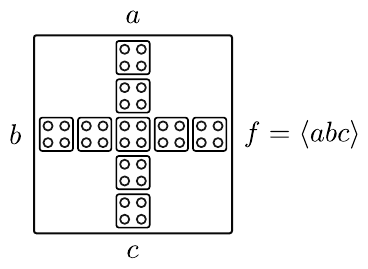}\label{fig:qca_gate}}\hfil
	\subfloat[SiDB OR gate.]{\quad\includegraphics[height=0.27\linewidth]{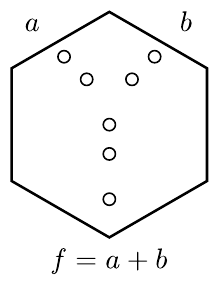}\quad\label{fig:sidb_gate}}
	\caption{FCN technology implementations.}
	\label{fig:basic_blocks}
\end{figure}

\begin{figure}[!t]
	\centering
	\subfloat[]{\includegraphics[width=0.175\linewidth]{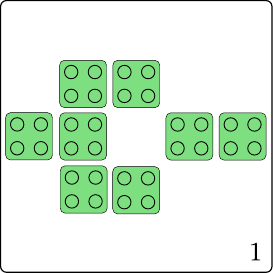}\label{fig:gates:inv}}\hfil
	\subfloat[]{\includegraphics[width=0.175\linewidth]{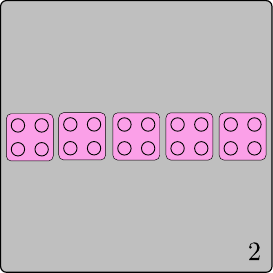}\label{fig:gates:wire}}\hfil
	\subfloat[]{\includegraphics[width=0.175\linewidth]{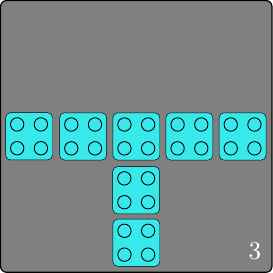}\label{fig:gates:splitter}}\hfil
	\subfloat[]{\includegraphics[width=0.175\linewidth]{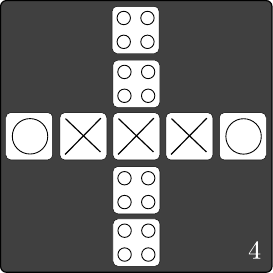}\label{fig:gates:3dcross}}\hfil
	\subfloat[]{\includegraphics[width=0.175\linewidth]{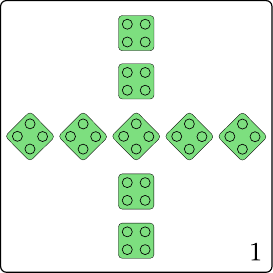}\label{fig:gates:cocross}}\hfil
	\caption{QCA ONE~\cite{reis2016methodology} gate implementations of an (a) Inverter, (b) wire, (c) fan-out, (d) 3D- and (e) co-planar wire crossing.}
	\label{fig:gates}
\end{figure}

\subsection{Cells and Gates}\label{sec:back:cell}
All FCN technologies share a fundamental building block called a \emph{cell}, which can represent binary encoding~\cite{Anderson14}. The logic state is determined by the cell's electrostatic polarization. When cells are placed in close proximity, they can polarize each other, transmitting signals without the need for electric current~\cite{lent2003molecular}.

A basic QCA cell has four quantum dots and two electrons, which, due to \emph{Coulomb} interactions, occupy opposite corners, resulting in two stable states for binary $0$ and $1$. SiDB cells follow a similar principle but use only two quantum dots. Both cell types and their states are illustrated in~\figurename~\ref{fig:cells}.

By exploiting the field interactions between adjacent cells, logic gates can be constructed using specific spatial arrangements of cells. QCA gates are typically arranged in squares, as illustrated in \figurename~\ref{fig:qca_gate}, while SiDB gates form hexagons, as shown in \figurename~\ref{fig:sidb_gate}. Numerous gates have been proposed for FCN technologies \cite{walter2022hexagons, tougaw1994logical, reis2016methodology}. In addition to the QCA majority gate depicted in \figurename~\ref{fig:qca_gate}, which can implement an AND or OR gate by fixing one input to $0$ or $1$, respectively, exemplary implementations of the QCA ONE~\cite{reis2016methodology} standard cell library are shown in \figurename~\ref{fig:gates}. An inverter can be achieved through diagonal coupling, as shown in \figurename~\ref{fig:gates:inv}. The gates shown in~\figurename~\ref{fig:gates:wire} and~\figurename~\ref{fig:gates:splitter} do not perform logic operations but instead serve as wire and fan-out implementations, respectively. This highlights that, in FCN, wires and gates have similar implications for fabrication costs, area usage, and implementation delays.

\subsection{Clocking}\label{sec:back:clock}

\begin{figure}[!tp]
	\centering
	\subfloat[2DDWave~\cite{vankamamidi2006clocking}.\label{fig:2DDwave}]{\includegraphics[height=0.13\textwidth]{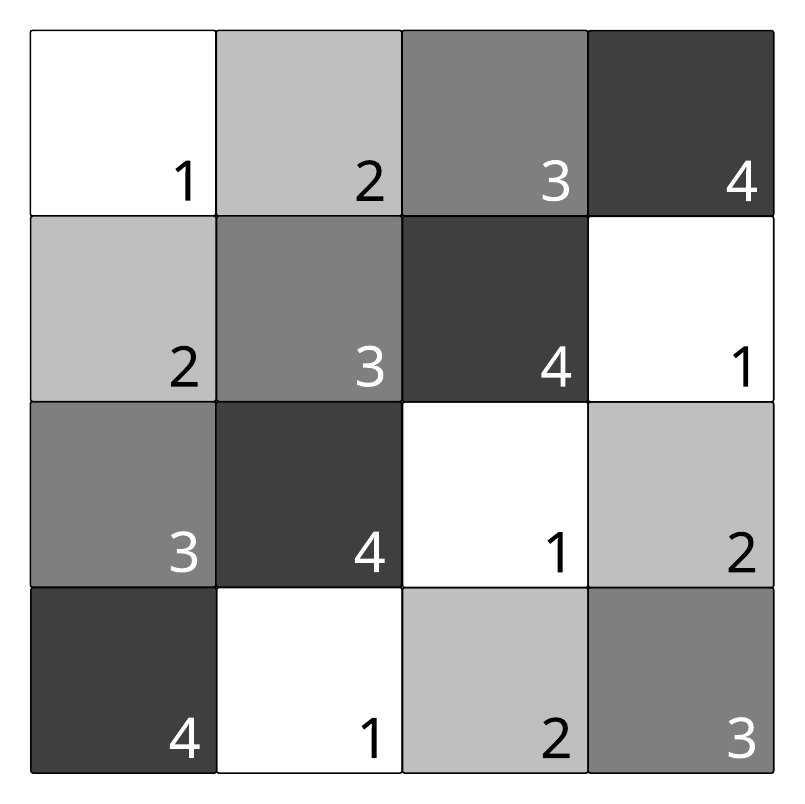}} \hfil
	\subfloat[USE~\cite{Campos16}.\label{fig:use}]{\includegraphics[height=0.13\textwidth]{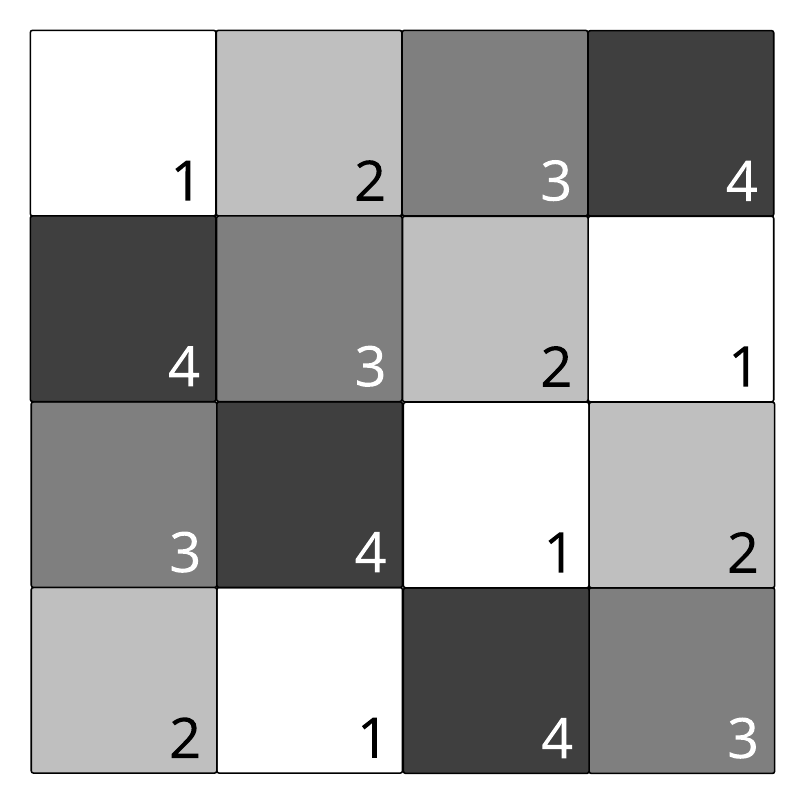}} \hfil
	\subfloat[RES~\cite{goswami2019efficient}.\label{fig:res}]{\includegraphics[height=0.13\textwidth]{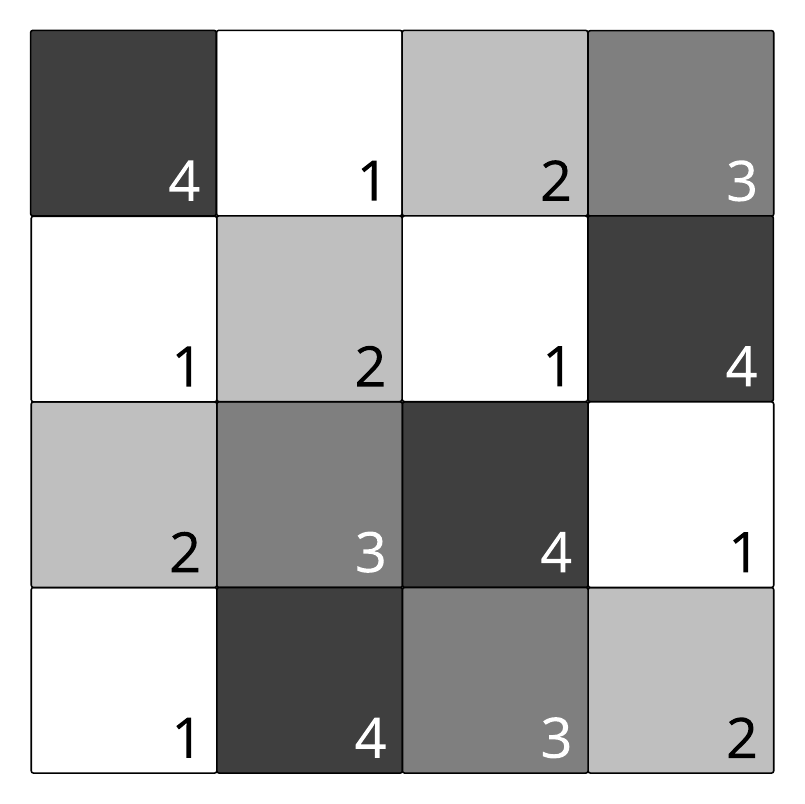}}
	\caption{Common clocking schemes for FCN circuit layouts.\label{fig:clocking}}
\end{figure}

FCN circuits utilize clocking to synchronize signal propagation and control information flow. To efficiently manage clocking in large circuits, FCN layouts are subdivided into uniform sections called \emph{tiles}, which are activated by an external field known as the \emph{clock}. This clock is distributed via buried electrodes in the substrate~\cite{Lent1997, Hennessy01}. The clock works in four phases, numbered 1 through 4, enabling a pipeline-like signal flow through the circuit. Each tile holds either a gate or a wire segment, as seen in~\figurename~\ref{fig:gates}. When constructing a layout, signal synchronization is essential. To achieve this, adjacent tiles must possess consecutive clock phases, referred to as \emph{local} synchronization, and signals that meet at the same tile must have traveled the same number of tiles, known as \emph{global} synchronization. The layout design process can be simplified by using predefined extensible clocking schemes and placing gates and wires accordingly. Popular clocking schemes from the literature are visualized in~\figurename~\ref{fig:clocking}.

\section{Related Work}\label{sec:related}

While state-of-the-art research has considered wire crossings feasible for FCN layouts~\cite{hofmann2024born, hofmann2023post}, recent studies show that their physical implementations for both QCA and SiDBs are not viable for functional circuits~\cite{retallick2017embeddingquantum, drewniok2023temperatureIEEE}.

Two QCA wire crossing implementations have been proposed~\cite{tougaw1994logical}. The first, 3D wire crossings (\figurename~\ref{fig:gates:3dcross}), use a second QCA layer, allowing one signal to move up while the other remains on the original plane before returning. However, current fabrication technologies only support quantum-dot generation on a single silicon surface~\cite{Wolkow14}, and multi-layer implementations are not physically feasible at present, rendering this approach impractical.

The second approach, co-planar crossings (\figurename~\ref{fig:gates:cocross}), rotates one wire by $45^\circ$ to avoid signal interference between the two orientations, enabling signals to cross on the same plane. However, quantum annealer simulations show a severe reliability drop, with logic correctness falling to $\SI{60}{\%}$\cite{retallick2017embeddingquantum}, making it unsuitable for dependable logic circuits. Furthermore, additional wiring is required to convert between rotated and standard cells, further reducing reliability and increasing area overhead, making this method equally impractical.

For SiDB crossings, simulations of various designs reveal extreme temperature sensitivity~\cite{drewniok2023temperatureIEEE}. The most robust implementation operates reliably only up to $\SI{21.78}{\kelvin}$, far below liquid nitrogen levels (~\SI{77}{\kelvin}), contradicting FCN’s energy efficiency goals and making such crossings infeasible.

These combined findings show that wire crossings do not function properly in FCN, necessitating their exclusion from the physical design of FCN. To the authors' knowledge, stateof-the-art algorithms rely on wire crossings and are therefore considered impractical in this work. To address this issue, this work presents a novel P\&R algorithm that eliminates wire crossings. FCN layouts without wire crossings are subsequently referred to as \emph{planar}.

\section{Proposed Approach}\label{sec:proposed}
\begin{figure}[!t]
	\centering
	\includegraphics[width=1.0\linewidth]{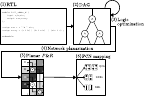}
	
	\caption{Proposed design flow for fully planar FCN circuit layouts, starting from a high-level description and eventually yielding a cell-level layout of a specific technology.}
	\label{fig:flow}
\end{figure}

This section presents the first scalable P\&R algorithm developed for fully planar FCN circuit layouts. To this end, first a high-level overview of a novel planar design flow for FCN is introduced in Section~\ref{sec:proposed:designflow}. From this high-level view, we explore the modifications needed at each step to  design planar FCN layouts. Achieving planarity at the logic synthesis level is detailed in Section~\ref{sec:proposed:preprocessing}. Section~\ref{sec:proposed:preservation} discusses how planarity is preserved throughout the transition from logic synthesis to the physical design. Finally, Section~\ref{sec:proposed:pr} presents the scalable P\&R algorithm, taking all these considerations into account.

\subsection{A Novel Design Flow for Fully Planar FCN Layouts}\label{sec:proposed:designflow}

The proposed design flow transforms an abstract circuit description into a planar FCN layout, systematically eliminating wire crossings at each stage. \figurename~\ref{fig:flow} illustrates this process.

It begins at the Register-Transfer Level (RTL) (1), where the circuit's logic is defined programmatically at a high level. This RTL description is converted into a logic network (2), typically represented as a Directed Acyclic Graph (DAG) using structures like And-Inverter Graphs (AIGs) for efficient optimization (3). Traditionally, area and delay are minimized by reducing node count and network depth~\cite{mishchenko2006dag}.

While area and delay remain important in FCN, minimizing wire crossings is crucial. At the logic network stage, some approaches reduce edge crossings as a proxy, even at the cost of area and delay~\cite{marakkalage2024technology, hien2024reducing}, while others eliminate crossings entirely to create planar logic networks~\cite{chaudhary2007fabricatable}. Planarizing the logic network (4) is a key preprocessing step in this work.

However, conventional P\&R algorithms do not preserve planarity, meaning that even if the input logic network is fully planar, wire crossings may still be introduced in the final layout. This fundamental limitation severely restricts the feasibility of FCN layouts.

To overcome this, we propose a P\&R algorithm that guarantees planarity throughout the process (5), ensuring a fully planar layout without reintroducing wire crossings. This marks a significant departure from prior methods, making large-scale FCN designs feasible. Finally, the FCN layout can be adapted across different implementations~\cite{hofmann2023hexagonalization}, and the cell representations of the gates are mapped onto the layout in the respective technology (6).

The next section details the first step---planarizing the logic network---as the foundation for a fully planar FCN layout.

\subsection{Network Planarization}\label{sec:proposed:preprocessing}
\begin{figure}[!t]
	\centering
	\subfloat[Input network.]{\includegraphics[height=0.34\linewidth]{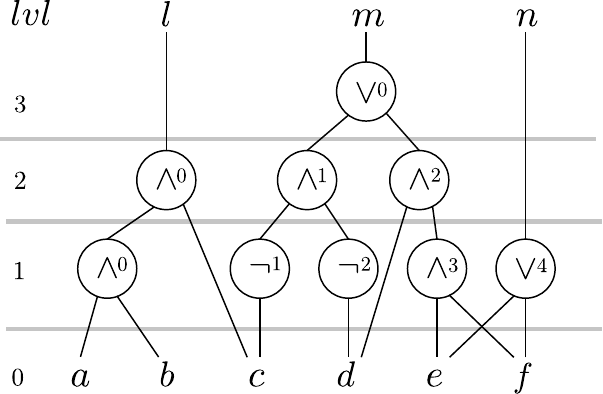}\label{fig:ntk:orig}}\hfil
	\subfloat[Planarized network.]{\includegraphics[height=0.34\linewidth]{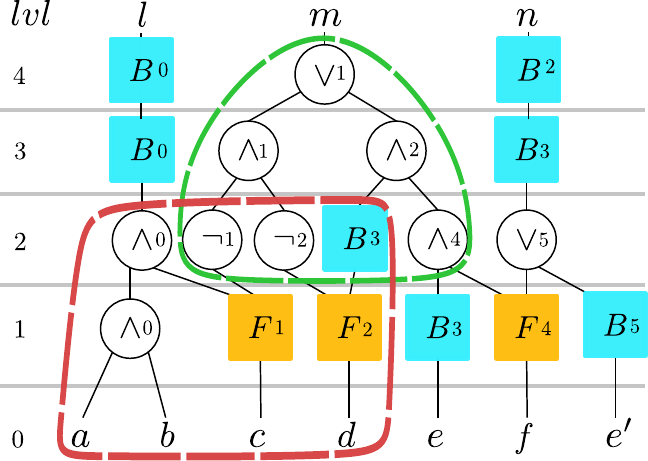}\label{fig:ntk:leg}}
	\caption{Logic network planarization through balancing and node duplication.}
	\label{fig:ntk}
\end{figure}

The planarization process for a logic network~\cite{chaudhary2007fabricatable} consists of three steps: fan-out substitution, network balancing, and node duplication. After fan-out substitution, only fan-out nodes have multiple outputs, typically limited to two signals, in accordance with FCN standard gate requirements~\cite{tougaw1994logical}.

The remaining steps prepare the network for a planar embedding---a drawing on a plane where no edges intersect. First, each node is assigned a level (the longest path from primary inputs (PIs) to the node) and a rank (its order within a level, numbered left to right). DAGs in this work are drawn bottom-up, with increasing levels. \figurename~\ref{fig:ntk:orig} illustrates an example network, showing levels on the left and ranks as small numbers beside each node’s function.

To achieve a fully planar embedding, edges spanning multiple levels must be segmented with buffers, ensuring each edge spans only one level—this step is called network balancing. Next, the node duplication algorithm iteratively processes levels from primary outputs (POs) to PIs, reordering and duplicating nodes to eliminate crossings between adjacent levels, ultimately producing a fully planar network~\cite{chaudhary2007fabricatable}.

\begin{example}
	Consider the logic network in~\figurename~\ref{fig:ntk:orig} and its planar embedding in~\figurename~\ref{fig:ntk:leg}. First, fan-out substitution introduces fan-out nodes (orange boxes). Next, network balancing inserts buffers (blue boxes) to divide multi-level edges. Finally, node duplication eliminates the crossing between level $1$ and the PIs by duplicating PI $e$. The assigned levels and ranks form a fully planar embedding of the network.
\end{example}

\subsection{Planarity Preservation}\label{sec:proposed:preservation}
The planar embedding generated in the previous step serves as input for P\&R. However, standard algorithms disregard this embedding, leading to wire crossings in the layout. To prevent this, the proposed algorithm must satisfy two key requirements: (1) the planar embedding must be preserved throughout P\&R, maintaining the nodes' levels and ranks in the final layout, and (2) signal transfer must follow the level-by-level structure ensured by the balancing step.

To achieve this in a tile-based layout, all nodes from a given network level are placed along a diagonal of tiles in rank order, ensuring the first requirement. The second requirement is met by assigning the same clock number to all tiles on a diagonal, enabling synchronized activation of nodes at that level. Adjacent levels are placed on consecutive diagonals, with ascending clock numbers matching uniform FCN clocking. This results in the 2DDWave scheme (\figurename~\ref{fig:2DDwave}), as illustrated by the following example:

\begin{example}
	A possible arrangement for the sub-network highlighted in green in~\figurename~\ref{fig:ntk:leg} is shown on the layout grid in~\figurename~\ref{fig:gaps:l1}. Level~2 nodes are placed along diagonal $d_1$, ordered by rank from top right to bottom left. All share clock number 1, represented by white tiles. The next level follows on diagonal $d_2$, assigned clock number 2, shown in light gray, and so forth.
\end{example}

Although placement alone provides a large solution space, the placement from the previous example is not straightforward. The nodes must be arranged to allow routing between adjacent levels. The combined P\&R problem is addressed in the following section.

\subsection{Placement \& Routing (P\&R)}\label{sec:proposed:pr}

\begin{figure}[!t]
	\centering
	\subfloat[Exemplary valid P\&R of green sub-network from ~\figurename~\ref{fig:ntk:leg}.]{\includegraphics[width=0.4\linewidth]{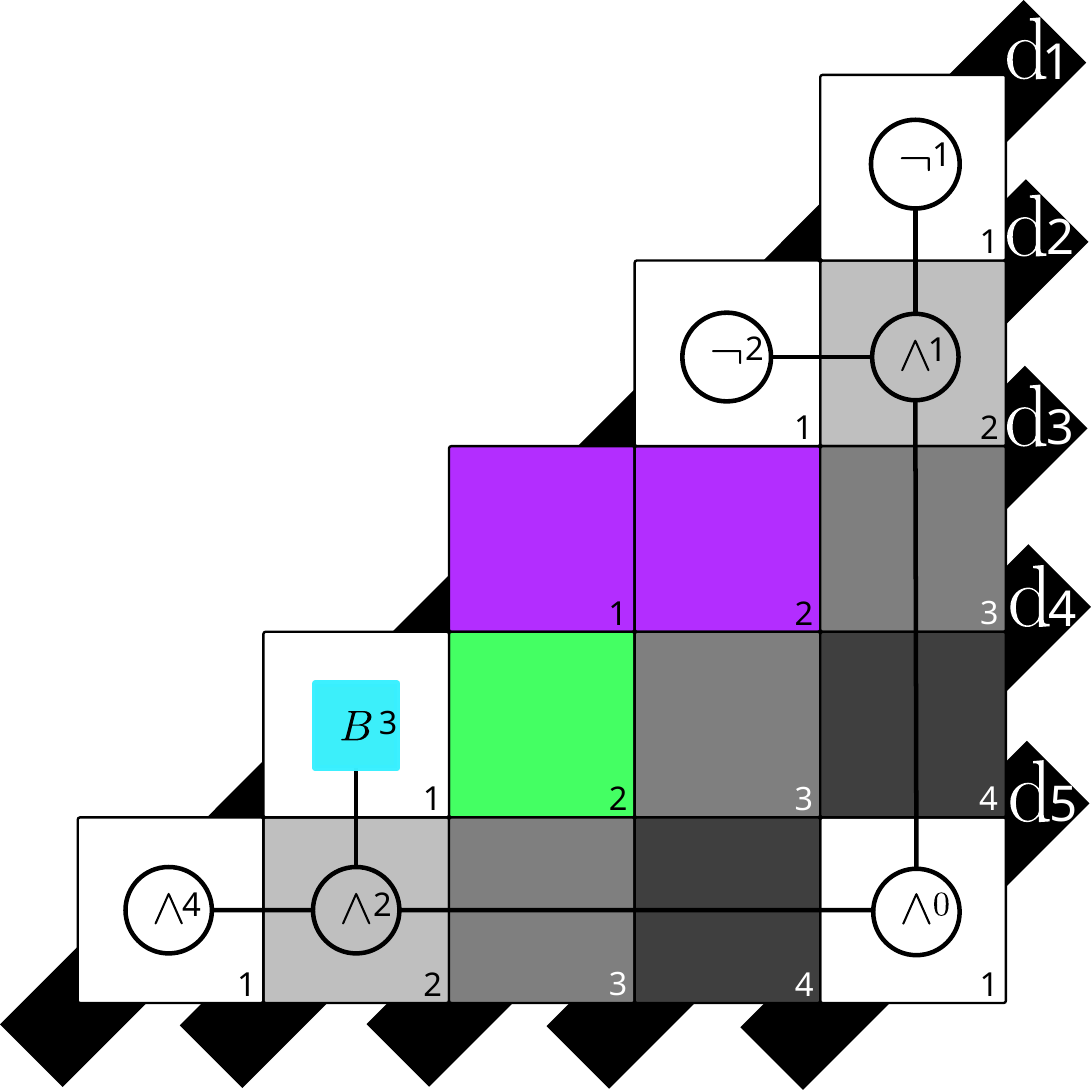}\label{fig:gaps:l1}} \hfil
	\subfloat[Routing issue due to invalid node placement of red sub-network from ~\figurename~\ref{fig:ntk:leg}.]{\includegraphics[width=0.4\linewidth]{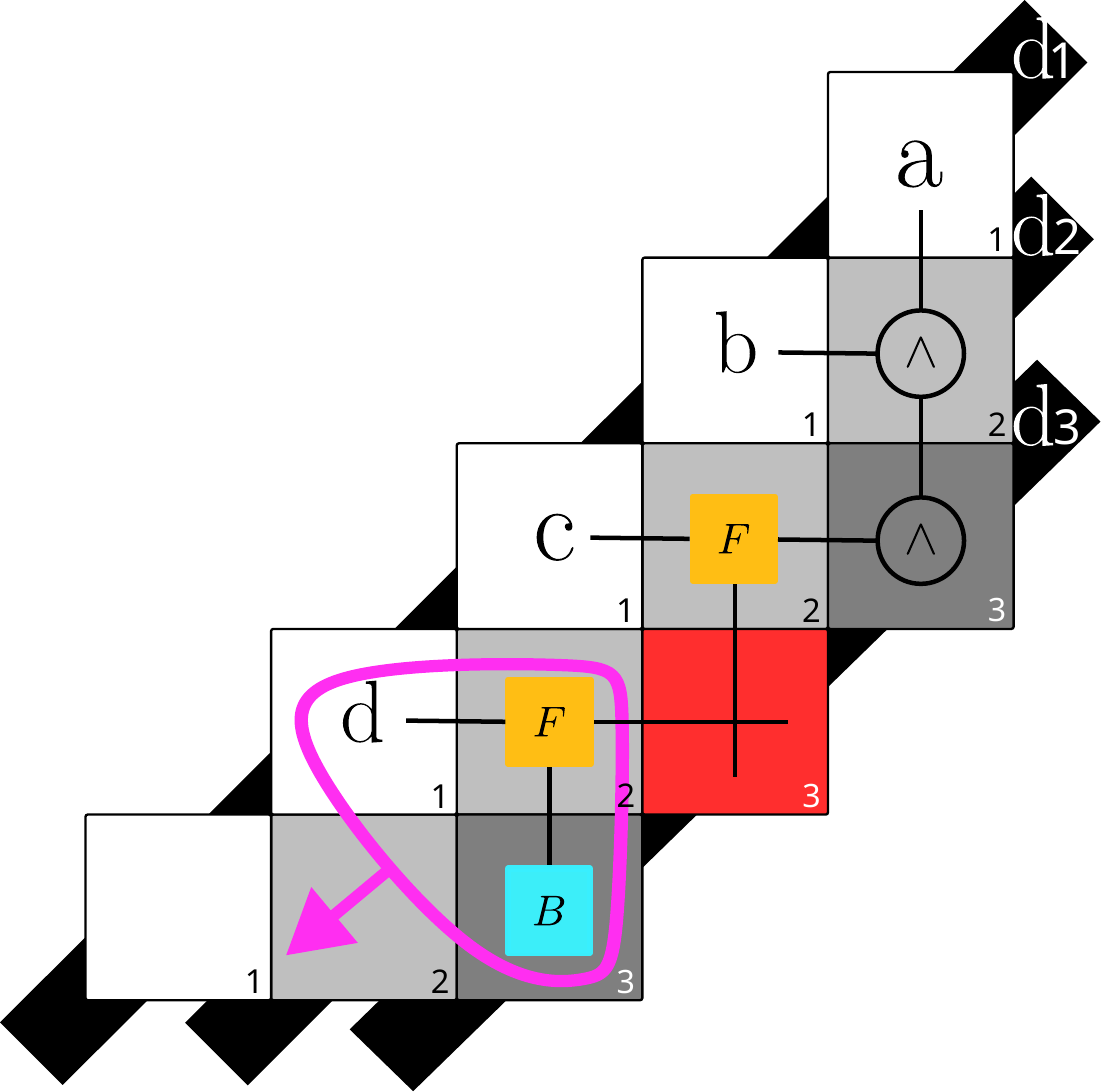}\label{fig:gaps:l2}}
	
	\caption{P\&R on 2DDWave clocking scheme.}
	\label{fig:gaps}
\end{figure}

The P\&R problem in FCN, inherently tied to clocking, is $\mathcal{N!P}$-complete~\cite{walter2019np}. However, our approach reduces the search space by predefining clocking and fixing relative node positions through the planar embedding. Additionally, network balancing restricts routing to adjacent levels, enabling iterative routing. This leads to the P\&R problem, described in the following.

With predefined clocking and fixed relative positions, a valid planar layout depends on two degrees of freedom. (1) Node placement: Nodes within a level are arranged diagonally by rank, but for valid routing to adjacent diagonals, they cannot always be placed directly next to each other. Empty tiles, referred to as \emph{gaps}, must be introduced, as shown in \figurename~\ref{fig:gaps:l1} (green and purple tiles). (2) Diagonal usage: Not all diagonals can be occupied, as seen in \figurename~\ref{fig:gaps:l1}, where diagonals $d_3$ and $d_4$ are left empty to enable proper routing.

We first define and compute gaps, which exist at each network level, corresponding to a diagonal in the layout. A gap occurs between adjacent nodes (i.e., nodes with consecutive ranks), with a gap value of 0 indicating direct adjacency. For a level $l$ with $N_l$ nodes, there are $N_l - 1$ gaps, stored in a gap vector $v_l$. Each level has its own gap vector, and all are combined into a gap array $A_G$ for the entire network.

Gap computation depends solely on the network structure, including levels, ranks, and fan-in/out relationships. It follows an iterative process, where gaps at each level are determined by local network structure and previously computed gaps in adjacent levels. The formation of gaps follows three distinct cases:

\begin{enumerate}
	\item \emph{2-ary gaps}: When two consecutive pairs of nodes feed into separate 2-ary nodes (e.g., AND gates) on the next diagonal, gaps arise due to placement constraints, as shown in~\figurename~\ref{fig:gaps:l1}. Each 2-ary node's position is determined by its predecessors: the northern predecessor sets the $x$-coordinate, and the western predecessor the $y$-coordinate ($\{x, y\} = \{x_{\text{north}}, y_{\text{west}}\}$). Since predecessors controlling the same coordinate (x or y) are spaced one tile apart, a gap (green tile) forms. The purple gap is addressed in the third case.
	
	\item \emph{Fan-out gaps}: Consider two adjacent fan-out nodes on the same diagonal, each with two distinct outputs. This requires placing four nodes on the next diagonal, but only three tiles are available for routing. As shown in~\figurename~\ref{fig:gaps:l2}, both fan-out nodes attempt to use the same red-marked tile, causing a conflict. Resolving this requires introducing a gap between the fan-out nodes.
	
	\item \emph{Propagated gaps}: Gaps can propagate \emph{forward} (PIs to POs) if an unconnected gap exists between two nodes at one level, as seen in~\figurename~\ref{fig:gaps:l1}, where a purple-highlighted gap on diagonal $d_1$ moves forward. Similarly, \emph{fan-out gaps} can propagate \emph{backward} (POs to PIs) if nodes remain unconnected in the previous level. In~\figurename~\ref{fig:gaps:l2}, pink-circled nodes shift, requiring both the fan-out node and its predecessor to adjust, affecting input spacing.
\end{enumerate}

Gap computation for individual levels, based on three cases, is integrated into an iterative algorithm that first computes gaps and then propagates them forward and backward to resolve routing conflicts. It begins at the PI level with forward iteration, computing gaps for each level. If a fan-out gap causes a conflict, the algorithm switches to backward iteration, adjusting earlier levels to resolve it. Forward iteration then resumes until the POs are reached, completing the gap array. In practice, conflicts are more frequent in lower levels, while higher levels typically have larger gaps, reducing conflicts.

As mentioned, the second degree of freedom in placement involves leaving empty diagonals. This is necessary when fan-ins of 2-ary nodes are separated by a gap, as shown in~\figurename~\ref{fig:gaps:l1}, where two AND nodes on diagonal $d_2$ with a gap of size 2 both feed into a node on diagonal $d_5$. For 2-ary nodes, where $\{x, y\} = \{x_{\text{north}}, y_{\text{west}}\}$, the number of empty diagonals $N_{d}$ matches the gap size between fan-ins. This computation is included in the algorithm, with the value inserted into the gap vector of the preceding level.

\begin{example}\label{ex:gap}
	In the planarized network (\figurename~\ref{fig:ntk:leg}), the algorithm starts with all gaps set to zero and propagates forward. At level $0$, no gaps form. At level $1$, two fan-out nodes lack a common fan-in at the next level, creating a fan-out gap. This gap propagates backward, adjusting the gap between PIs $c$ and $d$ in the preceding level. As a result, the updated gap vector for level $0$ is $v_0 = [0,0,1,0,0,0,0]$, indicating a gap of size 1 between PIs $c$ and $d$, with no gaps elsewhere. Forward propagation then resumes, iterating through all levels to compute the complete gap array:
	[[0,0,1,0,0,0,0],  
	[0,1,0,0,0,0],  
	[0,0,0,0,0,0],  
	[0,1,0,0],  
	[1,0,1]]

\end{example}

\begin{figure}[!t]
	\centering
	\includegraphics[height=0.64\linewidth]{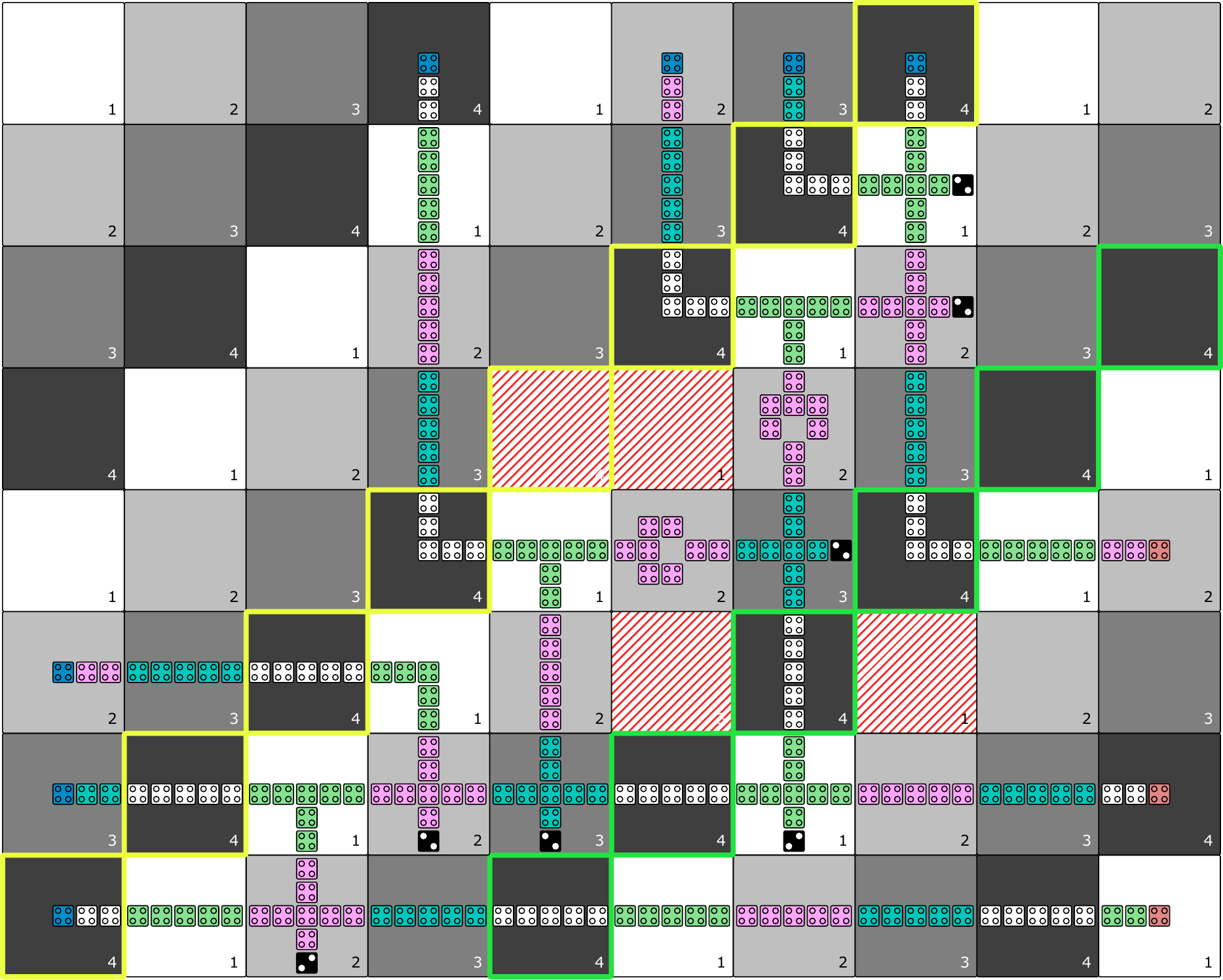}
	
	\caption{Planar layout of the network shown in~\figurename~\ref{fig:ntk:leg}. Yellow frames indicate the initial placement of PIs, red striped tiles represent gaps, and green-framed tiles mark an empty diagonal, following the algorithm used for scalable placement and routing.}
	\label{fig:lyt}
\end{figure}

After gap computation, the layout can be directly mapped, as illustrated in the following example:

\begin{example}
	\figurename~\ref{fig:lyt} shows the final layout for the network and gap array from Example~\ref{ex:gap}. The PIs are placed along a diagonal that accounts for their count and total gaps in the first level, highlighted by yellow frames, and are then routed to the layout's borders for accessibility.
	
	The gap between the second and third PI, represented in the first vector as [0,0,1,0,0,0,0], is marked with red stripes. Since the last entry of the next level’s vector [0,1,0,0,0,0] is zero, the next level is placed on the following diagonal. The same scheme applies to subsequent levels.
	
	A special case arises in the last level, where the final vector [1,0,1] ends with a 1, necessitating an empty diagonal, shown in green frames. This gap results from an AND node with predecessors separated by a gap of size 1. After all nodes are placed, the POs are routed to the layout’s borders.
\end{example}

\section{Experiments}\label{sec:experiments}

\begin{table*}[!t]
	\caption{Comparative experimental evaluation of the proposed fully planar physical design approach against the state of the art.}
	\centering
	\label{tab:benchmarks}
	\begin{minipage}{\linewidth}
		\centering
		\begin{adjustbox}{width=\linewidth}
			\begin{tabular}{
						@{}l@{\hskip 6pt} %
						@{}l@{\hskip 4pt} %
						S[table-format=5.0]@{\hskip 4pt} %
						S[table-format=2.0]@{\hskip 4pt} %
						S[table-format=6.0]@{\hskip 4pt} %
						r@{\hskip 4pt} %
						@{}l@{\hskip 4pt} %
						l@{\hskip 4pt} %
						S[table-format=3.2]@{\hskip 4pt} %
						r@{\hskip 4pt} %
						@{}l@{\hskip 4pt} %
						l@{\hskip 4pt} %
						S[table-format=2.2]@{\hskip 4pt} %
						S[table-format=3.2]@{\hskip 4pt} %
						S[table-format=6.0]@{\hskip 4pt} %
						r@{\hskip 4pt} %
						@{}l@{\hskip 4pt} %
						l@{\hskip 4pt} %
						S[table-format=2.2]@{\hskip 4pt} %
						r@{\hskip 4pt} %
						@{}l@{\hskip 4pt} %
						l@{\hskip 4pt} %
						S[table-format=2.2]@{\hskip 4pt} %
						S[table-format=3.2]@{\hskip 4pt} %
						S[table-format=3.2]@{\hskip 4pt} %
						S[table-format=4.2]@{\hskip 4pt} %
						S[table-format=3.2]@{\hskip 4pt} %
					}
\toprule
				                                 \multicolumn{5}{c}{\textsc{Benchmark Circuit}}                                  &                                                            \multicolumn{9}{c}{\textsc{SotA~\cite{hofmann2024born}}}                                                            &                                                            \multicolumn{10}{c}{\textsc{Proposed}}                                                             &                                       \multicolumn{3}{c}{\textsc{Difference}}                                         \\ %
				\cmidrule(lr){1-5} \cmidrule(lr){6-14} \cmidrule(lr){15-24} \cmidrule(lr){25-27}
				& Name            & $I$   & $O$ & $|N|$ & $W$ & $\times$ & $H$ & $t_{\text{gold}}[s]$ & $W_{\text{PLO}}$ & $\times$ & $H_{\text{PLO}}$ & $\Delta A_{\text{PLO}}[\%]$ & $t_{\text{PLO}}[s]$ & $|N|$  & $W$   & $\times$ & $H$   & $t_{\text{prop}}[s]$ & $W_{\text{PLO}}$ & $\times$ & $H_{\text{PLO}}$ & $\Delta A_{\text{PLO}}[\%]$ & $t_{\text{PLO}}[s]$ & $\Delta |N|[\%]$                     & $\Delta A[\%]$                        & $\Delta A_{\text{PLO}}[\%]$            \\ \midrule
				\multirow{7}{*}{\rotatebox[origin=c]{90}{Trindade~\cite{Trindade16}}}
				            & mux21           & 3     & 1   &   9   & 5   & $\times$ & 3   &                 0.10 & 5                & $\times$ & 3                & 0.00                        & 0.00                & 13     & 6     & $\times$ & 3     & 0.00                 & 5                & $\times$ & 3                & 16.67                       & 0.00                & 44.44                                & 20.00                                 & \pm0.00                               \\
				            & xor2            & 3     & 1   &   9   & 3   & $\times$ & 5   &                 0.09 & 3                & $\times$ & 5                & 0.00                        & 0.00                & 12     & 6     & $\times$ & 3     & 0.00                 & 5                & $\times$ & 3                & 16.67                       & 0.00                & 33.33                                & 20.00                                 & \pm0.00                               \\
				            & xnor2           & 4     & 1   &  11   & 3   & $\times$ & 6   &                 2.26 & 3                & $\times$ & 6                & 0.00                        & 0.00                & 17     & 8     & $\times$ & 4     & 0.00                 & 6                & $\times$ & 4                & 25.00                          & 0.00                & 54.55                                & 77.78                                 & 33.33                                 \\
				            & par\_gen        & 5     & 1   &  18   & 8   & $\times$ & 4   &                33.22 & 8                & $\times$ & 4                & 0.00                        & 0.00                & 33     & 11    & $\times$ & 6     & 0.00                 & 9                & $\times$ & 5                & 31.82                       & 0.00                & 83.33                                & 106.25                                & 40.63                                 \\
				            & HA              & 5     & 2   &  14   & 4   & $\times$ & 6   &                19.78 & 4                & $\times$ & 6                & 0.00                        & 0.00                & 26     & 8     & $\times$ & 5     & 0.00                 & 6                & $\times$ & 5                & 25.00                          & 0.00                & 85.71                                & 66.67                                 & 25.00                                 \\
				            & FA              & 7     & 2   &  16   & 6   & $\times$ & 4   &                19.19 & 6                & $\times$ & 4                & 0.00                        & 0.00                & 31     & 13    & $\times$ & 7     & 0.00                 & 5                & $\times$ & 7                & 61.54                       & 0.00                & 93.75                                & 279.17                                & 45.83                                 \\
				            & par\_check      & 16    & 1   &  42   & 5   & $\times$ & 17  &               100.00 & 5                & $\times$ & 17               & 0.00                        & 0.00                & 84     & 22    & $\times$ & 16    & 0.00                 & 10               & $\times$ & 17               & 51.70                        & 0.00                & 100.00                               & 314.12                                & 100.00                                \\ \midrule
				\multirow{16}{*}{\rotatebox[origin=c]{90}{Fontes~\cite{Fontes18}}}
				               & xor             & 4     & 1   &  10   & 5   & $\times$ & 3   &                 0.80 & 5                & $\times$ & 3                & 0.00                        & 0.00                & 16     & 7     & $\times$ & 4     & 0.00                 & 5                & $\times$ & 4                & 28.57                       & 0.00                & 60.00                                & 86.67                                 & 33.33                                 \\
				               & 1bitAdderAOIG   & 7     & 2   &  28   & 6   & $\times$ & 9   &               100.00 & 6                & $\times$ & 9                & 0.00                        & 0.00                & 52     & 15    & $\times$ & 7     & 0.00                 & 11               & $\times$ & 7                & 26.67                       & 0.00                & 85.71                                & 94.44                                 & 42.59                                 \\
				               & t\_5            & 7     & 2   &  21   & 6   & $\times$ & 8   &               100.00 & 6                & $\times$ & 8                & 0.00                        & 0.00                & 35     & 11    & $\times$ & 8     & 0.00                 & 8                & $\times$ & 8                & 27.27                       & 0.00                & 66.67                                & 83.33                                 & 33.33                                 \\
				               & t               & 8     & 2   &  23   & 10  & $\times$ & 4   &               100.00 & 10               & $\times$ & 4                & 0.00                        & 0.00                & 39     & 11    & $\times$ & 8     & 0.00                 & 10               & $\times$ & 8                & 9.09                        & 0.00                & 69.57                                & 120.00                                & 100.00                                \\
				               & c17             & 8     & 2   &  20   & 6   & $\times$ & 6   &               100.00 & 6                & $\times$ & 6                & 0.00                        & 0.00                & 41     & 15    & $\times$ & 8     & 0.00                 & 8                & $\times$ & 8                & 46.67                       & 0.00                & 105.00                               & 233.33                                & 77.78                                 \\
				               & b1\_r1          & 8     & 4   &  28   & 5   & $\times$ & 10  &               100.00 & 5                & $\times$ & 10               & 0.00                        & 0.00                & 50     & 11    & $\times$ & 9     & 0.00                 & 8                & $\times$ & 9                & 27.27                       & 0.00                & 78.57                                & 98.00                                 & 44.00                                 \\
				               & majority        & 8     & 1   &  27   & 8   & $\times$ & 12  &               100.00 & 8                & $\times$ & 12               & 0.00                        & 0.00                & 54     & 11    & $\times$ & 9     & 0.00                 & 10               & $\times$ & 9                & 9.09                        & 0.00                & 100.00                               & 3.13                                  & -6.25                                 \\
				               & newtag          & 9     & 1   &  28   & 6   & $\times$ & 10  &               100.00 & 6                & $\times$ & 10               & 0.00                        & 0.00                & 49     & 14    & $\times$ & 9     & 0.00                 & 11               & $\times$ & 8                & 30.16                       & 0.00                & 75.00                                & 110.00                                & 46.67                                 \\
				               & majority\_5\_r1 & 10    & 1   &  29   & 5   & $\times$ & 11  &               100.00 & 5                & $\times$ & 11               & 0.00                        & 0.00                & 52     & 14    & $\times$ & 10    & 0.00                 & 9                & $\times$ & 11               & 29.29                       & 0.00                & 79.31                                & 154.55                                & 80.00                                 \\
				               & 1bitAdderMaj    & 14    & 1   &  72   & 43  & $\times$ & 8   &               100.00 & 43               & $\times$ & 8                & 0.00                        & 0.00                & 151    & 44    & $\times$ & 22    & 0.00                 & 21               & $\times$ & 16               & 65.29                       & 0.00                & 109.72                               & 181.40                                & -2.33                                 \\
				               & xor5\_r1        & 16    & 1   &  70   & 9   & $\times$ & 24  &               100.00 & 9                & $\times$ & 24               & 0.00                        & 0.00                & 127    & 35    & $\times$ & 20    & 0.00                 & 19               & $\times$ & 17               & 53.86                       & 0.00                & 81.43                                & 224.07                                & 49.54                                 \\
				               & clpl            & 18    & 5   &  42   &                                                 \multicolumn{9}{c}{\emph{timeout limit reached}}                                                 & 158    & 50    & $\times$ & 19    & 0.00                 & 18               & $\times$ & 17               & 67.79                       & 0.01                & 276.19                               &                                       &                                       \\
				               & cm82a\_5        & 22    & 3   &  91   & 48  & $\times$ & 6   &               100.00 & 48               & $\times$ & 6                & 0.00                        & 0.00                & 186    & 44    & $\times$ & 27    & 0.00                 & 28               & $\times$ & 22               & 48.15                       & 0.01                & 104.40                               & 312.50                                & 113.89                                \\
				               & 2bitAdderMaj    & 57    & 2   &  243  &                                                 \multicolumn{9}{c}{\emph{timeout limit reached}}                                                 & 573    & 129   & $\times$ & 61    & 0.00                 & 61               & $\times$ & 58               & 55.04                       & 0.13                & 135.80                               &                                       &                                       \\
				               & xor5Maj         & 108   & 1   &  591  &                                                 \multicolumn{9}{c}{\emph{timeout limit reached}}                                                 & 1215   & 400   & $\times$ & 188   & 0.00                 & 132              & $\times$ & 156              & 72.62                       & 1.80                 & 105.58                               &                                       &                                       \\
				               & parity          & 121   & 1   &  580  & 295 & $\times$ & 9   &               100.00 & 295              & $\times$ & 9                & 0.00                        & 0.07                & 820    & 209   & $\times$ & 162   & 0.00                 & 164              & $\times$ & 122              & 40.91                       & 0.45                & 41.38                                & 1175.25                               & 653.60                                \\ \midrule
				\multirow{10}{*}{\rotatebox[origin=c]{90}{IWLS93~\cite{IWLS93}}}
				                 & x4              & 1749  & 71  & 3413  &                                                 \multicolumn{9}{c}{\emph{timeout limit reached}}                                                 & 20960  & 2470  & $\times$ & 1888  & 0.27                 & 2362             & $\times$ & 1848             & 6.40                         & 104.42              &                                      &                                       &                                       \\
				                 & duke2           & 2109  & 29  & 3737  &                                                 \multicolumn{9}{c}{\emph{timeout limit reached}}                                                 & 20970  & 2775  & $\times$ & 2131  & 0.33                 & 2739             & $\times$ & 2116             & 1.99                        & 102.50              &                                      &                                       &                                       \\
				                 & rd84            & 3239  & 4   & 5196  &                                                 \multicolumn{9}{c}{\emph{timeout limit reached}}                                                 & 24899  & 4031  & $\times$ & 3241  & 0.78                 & 4016             & $\times$ & 3237             & 0.50                         & 111.46              &                                      &                                       &                                       \\
				                 & t481            & 4208  & 1   & 7144  &                                                 \multicolumn{9}{c}{\emph{timeout limit reached}}                                                 & 44244  & 8145  & $\times$ & 4213  & 1.52                 & 8134             & $\times$ & 4211             & 0.18                        & 123.92              &                                      &                                       &                                       \\
				                 & c880            & 4737  & 26  & 8296  &                                                 \multicolumn{9}{c}{\emph{timeout limit reached}}                                                 & 44266  & 13365 & $\times$ & 4765  & 2.23                 & 13348            & $\times$ & 4754             & 0.36                        & 132.85              &                                      &                                       &                                       \\
				                 & vda             & 5235  & 39  & 8679  &                                                 \multicolumn{9}{c}{\emph{timeout limit reached}}                                                 & 66960  & 6171  & $\times$ & 5303  & 3.99                 & 6162             & $\times$ & 5294             & 0.32                        & 141.97              &                                      &                                       &                                       \\
				                 & table5          & 5235  & 39  & 11255 &                                                 \multicolumn{9}{c}{\emph{timeout limit reached}}                                                 & 62693  & 6171  & $\times$ & 5303  & 3.99                 & 6162             & $\times$ & 5294             & 0.32                        & 141.97              &                                      &                                       &                                       \\
				                 & table3          & 6988  & 14  & 11261 &                                                 \multicolumn{9}{c}{\emph{timeout limit reached}}                                                 & 68878  & 11030 & $\times$ & 6993  & 5.47                 & 11030            & $\times$ & 6992             & 0.01                        & 185.50              &                                      &                                       &                                       \\
				                 & apex3           & 8071  & 50  & 13633 &                                                 \multicolumn{9}{c}{\emph{timeout limit reached}}                                                 & 91970  & 11422 & $\times$ & 8116  & 8.09                 & 11422            & $\times$ & 8109             & 0.09                        & 238.79              &                                      &                                       &                                       \\
				                 & cordic          & 15148 & 2   & 25104 &                                                 \multicolumn{9}{c}{\emph{timeout limit reached}}                                                 & 149182 & 41884 & $\times$ & 15157 & 31.85                & 41884            & $\times$ & 15157            & 0.00                        & 758.13              &                                      &                                       &                                       \\ \midrule
				\multicolumn{24}{l}{\emph{Average}}                                                                                                                                                                                                                                                                                                                                                                                                         & \multicolumn{1}{r}{$\mathbf{77.59}$} & \multicolumn{1}{r}{$\mathbf{188.03}$} & \multicolumn{1}{r}{$\mathbf{75.55}$}  \\ \bottomrule
				\vspace{-0.1em}
			\end{tabular}
		\end{adjustbox}
		\begin{minipage}{\linewidth}
			\scriptsize	
			For each benchmark, $I$ and $O$ represent the logic network's number of primary inputs (PIs) and outputs (POs), respectively. The number of nodes (including PIs and POs) is given as $|N|$. For the gold algorithm, the buffers introduced during planarization are removed, resulting in a smaller node count. The layout dimensions, represented as width ($W$) and height ($H$), are provided both before and after post-layout optimization (PLO). The area improvement from PLO for each algorithm is denoted as $\Delta A_{\text{PLO}}$. Run times for the $P \& R$ algorithms are given as $t_{\text{prop}}$, $t_{\text{gold}}$, and for PLO as $t_{\text{PLO}}$. The differences between the two algorithms are tracked in terms of the number of nodes in the logic network (due to removed buffers), denoted as $\Delta |N|$, as well as the area difference before PLO ($\Delta A$) and after PLO ($\Delta A_{\text{PLO}}$).
		\end{minipage}
	\end{minipage}
\end{table*}

The experiments were conducted on an AMD Ryzen 7 PRO 6850U with \SI{32}{\giga\byte} of DDR5 RAM. To promote open research and data sharing, both the planarization and scalable P\&R algorithms are publicly available as part of the \emph{Munich Nanotech Toolkit} (MNT) at \url{https://github.com/cda-tum/fiction}.

The state-of-the-art \emph{graph-oriented layout design} (\emph{gold}, \cite{hofmann2024born}) was modified to generate planar layouts and executed with a $\SI{100}{\second}$ timeout, serving as the reference. Both \emph{gold} and the proposed algorithm were tested on benchmarks from \emph{Trindade}~\cite{Trindade16}, \emph{Fontes}~\cite{Fontes18}, and \emph{IWLS93}~\cite{IWLS93}. A post-layout optimization (PLO) algorithm~\cite{hofmann2023post, hofmann2024wiring} was adapted to maintain planarity and applied to layouts from both methods, minimizing area by rearranging gates. Logical correctness was verified using formal methods~\cite{walter2020verify}.

Table~\ref{tab:benchmarks} presents the results. The left-most column lists each benchmark with its number of primary inputs ($I$) and outputs ($O$). The initial node count in the input logic network is provided for both approaches. Execution times for P\&R and PLO, optimized layout area, and area improvement are also reported. PLO was executed with a $\SI{100}{\second}$ soft timeout. The \textsc{Difference} column tracks node count differences due to buffer addition/removal, area after P\&R, and area after PLO. The final row averages results over benchmarks where \emph{gold} produced a valid layout.

For the \emph{Trindade} and \emph{Fontes} suites, input networks for the proposed algorithm contain $\SI{77.59}{\percent}$ more nodes on average, leading to a $\SI{188.03}{\percent}$ area overhead. PLO reduces this to $\SI{75.55}{\percent}$ but finds no improvements for \emph{gold} layouts, which are already near-optimal~\cite{hofmann2024born}. Despite this, the proposed algorithm completes all benchmarks in under $\SI{0.01}{\second}$, whereas \emph{gold} times out on \emph{clpl}, \emph{2bitAdderMaj}, and \emph{xor5Maj}.

Scalability is further demonstrated by select \emph{IWLS} benchmarks~\cite{IWLS93}, where node counts range from $20\mathrm{k}$ to $150\mathrm{k}$. The proposed approach successfully processes all benchmarks, while \emph{gold} times out on each. The largest benchmark handled, \emph{cordic}, is $182\times$ larger than \emph{parity}, the largest benchmark \emph{gold} could process. Due to PLO's scalability limitations, optimization improvements decline with increasing benchmark size, from $\SI{6.4}{\percent}$ for \emph{x4} to $\SI{0.0}{\percent}$ for \emph{cordic}.

These results demonstrate that the proposed algorithm scales efficiently for large benchmarks, while the adapted \emph{gold} algorithm performs better on smaller circuits. Real-world circuits often contain thousands of nodes, making scalability crucial, especially for planar FCN layouts, where ensuring planarity increases the node count due to duplication. Thus, a scalable approach is essential for practical planar FCN design.

\section{Conclusion}\label{sec:conclusion}

This work introduced a scalable placement and routing~(P\&R) algorithm for Field-Coupled Nanocomputing (FCN), eliminating wire crossings through a fully planar design flow. By addressing the impracticality of wire crossings, we removed a key barrier in FCN's path to scalability and established a framework that aligns with FCN's distinctive layout constraints. Our approach extends prior methods, scaling beyond small networks to handle circuits up to $182 \times$ larger, marking a crucial step toward large-scale, real-world FCN applications.

Future work will focus on refining optimization techniques across the design flow, further improving area efficiency and computational performance. These enhancements will enable FCN to better address the demands of modern computing and support even larger, more complex circuits.

	\balance
	\bibliographystyle{IEEEtran}
	\bibliography{./bib/IEEEabrv, ./bib/Bibliography.bib}
\end{document}